\documentclass[proof]{WileyASNA-v1}

\usepackage{graphicx}

\articletype{Original Article}%

\received{8 September 2019}
\revised{31 October 2019}
\accepted{31 October 2019}

\begin{document}

\title{How do atomic code uncertainties affect abundance measurements in the intracluster medium?}

\author[1,2,3]{F. Mernier*}

\author[1,4,5]{N. Werner}

\author[1]{K. Lakhchaura}

\author[3]{J. de Plaa}

\author[6]{L. Gu}

\author[3,7]{J. S. Kaastra}

\author[8]{J. Mao}

\author[3,7,9]{A. Simionescu}

\author[3,7]{I. Urdampilleta}

\authormark{F. MERNIER \textsc{et al}}

\address[1]{\orgname{MTA-E\"otv\"os University Lend\"ulet Hot Universe Research Group}, \orgaddress{\city{Budapest}, \country{Hungary}}}

\address[2]{\orgdiv{Institute of Physics}, \orgname{E\"otv\"os University}, \orgaddress{\city{Budapest}, \country{Hungary}}}

\address[3]{\orgname{SRON Netherlands Institute for Space Research}, \orgaddress{\city{Utrecht}, \country{The Netherlands}}}

\address[4]{\orgdiv{Department of Theoretical Physics and Astrophysics, Faculty of Science}, \orgname{Masaryk University}, \orgaddress{\city{Brno}, \country{Czech Republic}}}

\address[5]{\orgdiv{School of Science}, \orgname{Hiroshima University}, \orgaddress{\city{Higashi-Hiroshima}, \country{Japan}}}

\address[6]{\orgname{RIKEN High Energy Astrophysics Laboratory}, \orgaddress{\city{Saitama}, \country{Japan}}}

\address[7]{\orgdiv{Leiden Observatory}, \orgname{Leiden University}, \orgaddress{\city{Leiden}, \country{The Netherlands}}}

\address[8]{\orgdiv{Department of Physics}, \orgname{University of Strathclyde}, \orgaddress{\city{Glasgow}, \country{UK}}}

\address[9]{\orgdiv{Kavli Institute for the Physics and Mathematics of the Universe (WPI)}, \orgname{University of Tokyo}, \orgaddress{\city{Kashiwa}, \country{Japan}}}

\corres{*F. Mernier, MTA-E\"otv\"os University Lend\"ulet Hot Universe Research Group, P\'azm\'any P\'eter s\'et\'any 1/A, Budapest, 1117, Hungary. \email{mernier@caesar.elte.hu}}


\abstract{Accurate chemical abundance measurements of X-ray emitting atmospheres pervading massive galaxies, galaxy groups, and clusters provide essential information on the star formation and chemical enrichment histories of these large scale structures. Although the collisionally ionised nature of the intracluster medium (ICM) makes these abundance measurements relatively easy to derive, underlying spectral models can rely on different atomic codes, which brings additional uncertainties on the inferred abundances. Here, we provide a simple, yet comprehensive comparison between the codes SPEXACT v3.0.5 (\texttt{cie} model) and AtomDB v3.0.9 (\texttt{vapec} model) in the case of moderate, CCD-like resolution spectroscopy. We show that, in cool plasmas ($kT \lesssim 2$ keV), systematic differences up to $\sim$20\% for the Fe abundance and $\sim$45\% for the O/Fe, Mg/Fe, Si/Fe, and S/Fe ratios may still occur. Importantly, these discrepancies are also found to be instrument-dependent, at least for the absolute Fe abundance. Future improvements in these two codes will be necessary to better address questions on the ICM enrichment.}


\jnlcitation{\cname{%
\author{F. Mernier}, 
\author{N. Werner}, 
\author{K. Lakhchaura}, 
\author{J. de Plaa}, 
\author{L. Gu}, 
\author{J. S. Kaastra}, 
\author{J. Mao}, 
\author{A. Simionescu}, 
and \author{I. Urdampilleta}} (\cyear{2019}), 
\ctitle{How do atomic code uncertainties affect the abundance measurements of the intracluster medium?}, \cjournal{Astron. Nachr. /AN.}, \cvol{2019;00:1--6}.}


\maketitle

\footnotetext{\textbf{Abbreviations:} ICM, intracluster medium; CIE, collisional ionisation equilibrium; SNcc, core-collapse supernovae; SNIa, Type Ia supernovae}

\section{Metals in the intracluster medium}\label{sec:intro}

Being by essence the building blocks of interstellar molecules, dust, rocky planets, and even life, metals play a fundamental role in shaping the remarkable diversity of our Universe. As opposed to hydrogen and helium---the bulk of which have been synthesised a few minutes after the Big Bang, these heavier chemical elements find their origin in stars, and particularly at the end of their lifetimes \citep[for a review, see][]{nomoto2013}. While $\alpha$-elements (e.g. O, Ne, Mg) are mainly produced by the explosion of massive stars in a form of core-collapse supernovae (SNcc), heavier metals (e.g. Ca, Cr, Mn, Fe, Ni) mainly originate from Type Ia supernovae (SNIa), after a white dwarf in a binary system burns its carbon in an explosive way \citep[e.g.][]{nomoto2018,thielemann2018}. Intermediate mass elements (e.g. Si, Si, Ar) are produced by SNcc and SNIa in comparable amounts. Finally, lighter metals such as C and N are thought to be produced in low-mass stars, during their asymptotic giant branch (AGB) phase \citep[e.g.][]{karakas2010}.

Not only these freshly created metals were able to enrich their immediate surroundings and help forming new stars, but they could also partly escape out of the gravitational well of their host galaxies. In fact, the presence of emission lines in the X-ray spectra of the hot ($10^6$--$10^8$ K), highly ionised atmospheres surrounding the most massive galaxies and pervading galaxy groups and clusters is the smoking gun evidence that chemical enrichment is at play even within these large scale structures \citep[e.g.][]{mitchell1976,lea1982}. The presence of metals in the intracluster medium (ICM) naturally poses several fundamental questions \citep[for recent reviews, see][]{biffi2018c,mernier2018c}, among which: when (and how) did the ICM get enriched? The key to answer this question resides in the overall evolution of the ICM metallicity with cosmic time. Despite the impressive efforts that have been dedicated to this aspect so far \citep[e.g.][]{ettori2015,mcdonald2016,mantz2017,liu2018} the limited collecting area of current X-ray missions (e.g. \textit{XMM-Newton}, \textit{Chandra}, \textit{Suzaku}) translates into difficulties of quantifying accurately the chemical evolution of the ICM. Alternatively, and interestingly, remarkable signatures of the past chemical history of \textit{nearby} clusters and groups can be found in the spatial distribution of their metals. The clearest example is arguably the uniform metallicity profile measured towards cluster outskirts (i.e. beyond $\sim$0.5 $r_{500}$\footnote{By convention, $r_{500}$ delimitates the radius within which the mean cluster gas density reaches 500 times the critical density of the Universe.}) as an indirect evidence of an \textit{early} enrichment scenario, in which the bulk of metals were ejected outside galaxies and well mixed in the intergalactic space before clusters started to assemble \citep[][]{fujita2008,werner2013,urban2017}. These observations, along with this scenario, are in excellent agreement with cosmological simulations including early feedback from active galactic nuclei \citep[][]{biffi2017,biffi2018}. Central metal peaks typically seen in nearby cool-core systems also provide valuable information about clusters and groups chemical histories. For instance, the presence of such a peak in both Fe and $\alpha$-elements strongly suggests that these metals have little to do with the current "red-and-dead" stellar population of the central dominant galaxy \citep[][]{deplaa2006,simionescu2009,mernier2017}.

The low density of the ICM (translating into a negligible optical depth) coupled with its collisional ionisation equilibrium (CIE) makes its emission spectra relatively simple to model in terms of density, temperature, and chemical abundances. In particular, even using moderate resolution spectroscopy instruments, abundances can be measured more precisely in the ICM than in our own Solar System \citep[e.g.][]{deplaa2007,mernier2016a}. On paper, these ICM abundance ratios are invaluable because, as witnesses of billions of supernovae explosions, they can be directly compared to SNIa and SNcc yields expected from nucleosynthesis models and thus help to (dis)favour some of them \citep[e.g.][]{deplaa2007,mernier2016b,simionescu2019}. Whereas this exercise is, in practice, still difficult given the uncertainties related to the nucleosynthesis models themselves \citep{degrandi2009,mernier2016b,simionescu2019}, a clear picture that recently emerged---notably thanks to the exquisite spectral resolution provided by the \textit{Hitomi} observatory on the Perseus cluster---is that the ICM chemical composition is surprisingly similar to that of our own Solar System \citep{THC2017,mernier2018b,simionescu2019}. One notable exception to this trend is the significantly super-solar N/Fe abundance ratio measured in hot atmospheres of nearby clusters and groups, suggesting that AGB stars do contribute to the central ICM enrichment as well \citep[e.g.][]{werner2006,sanders2011,mao2019}.

\section{Atomic codes and systematic uncertainties}\label{sec:atomic_codes}

Precise measurements do not necessarily mean accurate measurements. This is particularly true for the routinely measured abundances which, despite the relatively simple physical properties of the ICM, may be affected by several sources of systematic biases, hence uncertainties. Among them, one can cite e.g. the potentially complex multi-temperature structure of the gas, the imperfect calibration of the instrumental response, or even background-related uncertainties \citep[for a detailed list of the well known systematic uncertainties that may affect the ICM abundances, see][]{mernier2018c}. 

Another (yet no less important) source of systematic uncertainties concerns our current knowledge of the atomic processes that produce the continuum and the emission lines in ICM spectra. For instance, it has been shown that improvements in atomic codes can significantly affect measurements of absolute Fe abundances in groups and ellipticals \citep{mernier2018a} and of X/Fe abundance ratios in more massive systems \citep{mernier2018b}, thereby altering their astrophysical interpretations. Nowadays, most of the ICM abundances reported in the literature rely on two sets of atomic codes/tables\footnote{In addition to these two codes, although less often used in the literature to fit X-ray spectra, one can also cite CHIANTI \citep{landi2013} and Cloudy \citep{ferland2017}.} (Table~\ref{table:list_codes}).

\begin{itemize}
\item SPEXACT (SPEX Atomic Code and Tables), which is a major update of the (now deprecated) \texttt{mekal} code \citep{mewe1985,mewe1986}. Since 1995, SPEXACT is available via the \texttt{cie} model in the SPEX fitting package\footnote{\href{https://www.sron.nl/astrophysics-spex}{https://www.sron.nl/astrophysics-spex}} \citep{kaastra1996,kaastra2018}. 
\item AtomDB, which is a database that has been continuously updated since the first code of \citet{raymond1977}. It is now implemented as the \texttt{apec} model (or the variant \texttt{vapec} to model the abundances individually) in the fitting package XSPEC\footnote{\href{https://heasarc.gsfc.nasa.gov/xanadu/xspec}{https://heasarc.gsfc.nasa.gov/xanadu/xspec}} \citep{smith2001,foster2012}.
\end{itemize}

\begin{center}
\begin{table}[b]%
\caption{\enspace List of the two plasma codes (and associated nomenclatures) considered in this work.\label{table:list_codes}}
\centering
\begin{tabular*}{0.5\textwidth}{@{\extracolsep\fill}lccccc}
\toprule
\textbf{Fitting} & \textbf{Plasma}  & \textbf{Atomic}  & \textbf{Current}  & \textbf{Ref.}   \\
\textbf{package} & \textbf{model}  & \textbf{code/tables}  & \textbf{version}  &    \\
\midrule
SPEX & \texttt{cie}  & SPEXACT  & 3.0.5  & \tnote{(1)}, \tnote{(2)}   \\
XSPEC & \texttt{(v)apec}  & AtomDB  & 3.0.9  & \tnote{(3)}, \tnote{(4)}   \\
\bottomrule
\end{tabular*}
\begin{tablenotes}
\item (1) \citet{kaastra1996}; (2) \citet{kaastra2018}; (3) \citet{smith2001}; (4) \citet{foster2012}
\end{tablenotes}
\end{table}
\end{center}

During their histories, these two codes have evolved independently, as they have used different atomic databases, approximations on the considered radiative processes, and methods for computing spectral models (i.e. calculated "on the spot" for \texttt{cie} vs. pre-calculated tables for \texttt{apec}/\texttt{vapec}). Since these two codes are not easily comparable as they are implemented in distinct fitting packages, many authors chose to rely on only one model to measure ICM temperatures or abundances. If the statistical errors of these best-fit parameters are small, the results may be affected by the choice of the code. On a more optimistic side, comparing the results predicted by these two independent codes constitutes an unique opportunity to first test, then improve our overall understanding of plasma emission processes. In this respect, the very high energy resolution spectrum of the Perseus cluster provided by the SXS instrument onboard \textit{Hitomi} allowed considerable improvements of both SPEXACT and AtomDB (respectively up to v3.0.3 and v3.0.8), thereby making them converge better than their previous versions before the launch of the mission \citep{THC2018_atomic}. Specifically, for an ICM of moderately hot temperature ($kT \sim 4$ keV), at SXS energy resolution ($\sim$5 eV) and energy range ($\sim$2--10 keV), discrepancies in the absolute abundances of Fe and other elements are now limited to $\sim$16\% and less than $\sim$11\%, respectively.

This relatively good agreement is certainly promising for future missions (e.g. \textit{XRISM}, \textit{Athena}). However, it should be kept in mind that (i) \textit{Hitomi} could not access the Fe-L complex of Perseus, in which the plethora of transitions would have probably revealed more code-related discrepancies to reduce; and (ii) even after the expected launch of \textit{XRISM} ($\sim$2021), the large majority of archival ICM spectra will remain at moderate energy resolution. Therefore, a systematic comparison between the most recent versions of these atomic codes (i.e. SPEXACT v3.0.5 and AtomDB v3.0.9; see Table \ref{table:list_codes}) at CCD-like resolution and within the full energy window of currently flying X-ray observatories (e.g. \textit{XMM-Newton}/EPIC, \textit{Chandra}/ACIS, eROSITA) is necessary to better quantify their expected systematic uncertainties on measured abundances.

\begin{figure*}[h]
\centering
\includegraphics[width=0.477\textwidth,trim={2.4cm 0 0 0},clip]{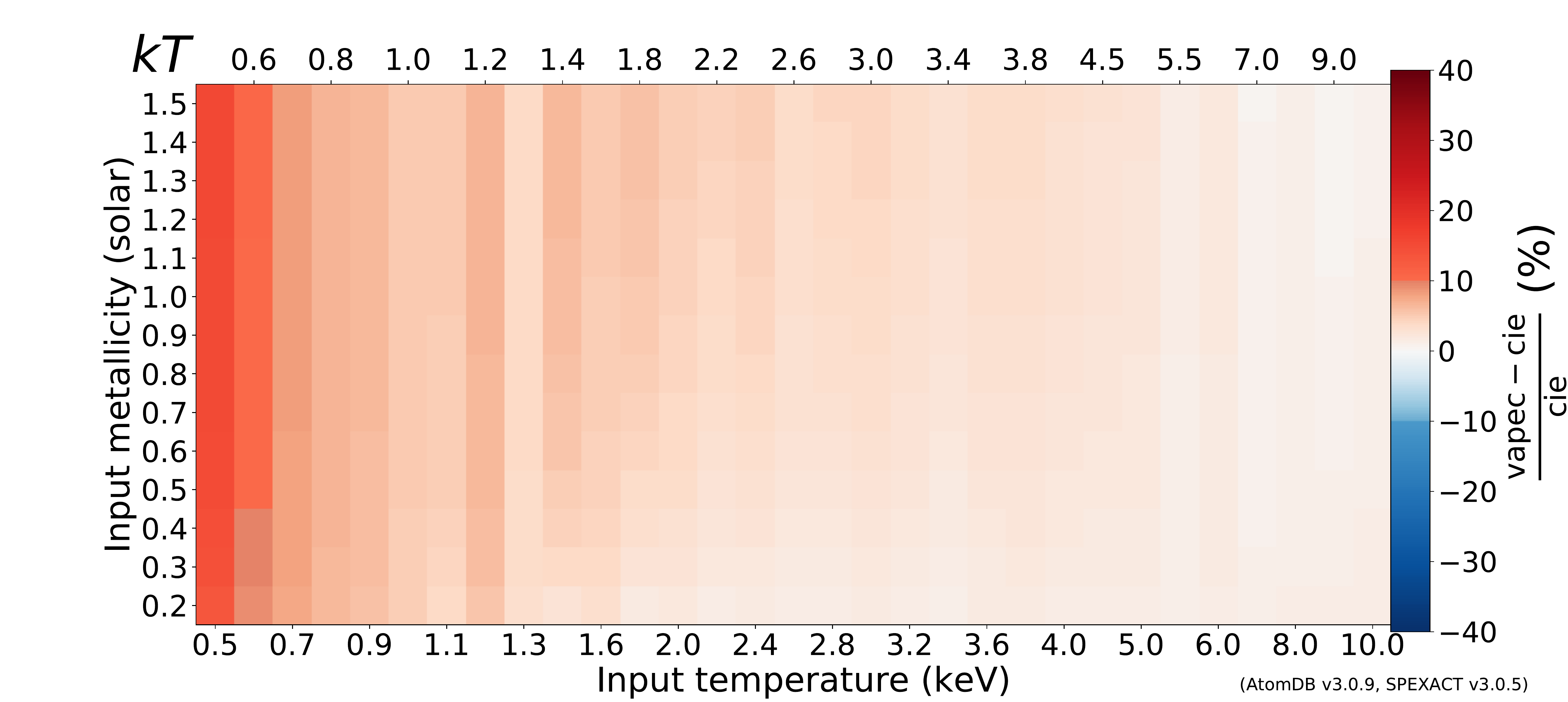}
\includegraphics[width=0.477\textwidth,trim={2.4cm 0 0 0},clip]{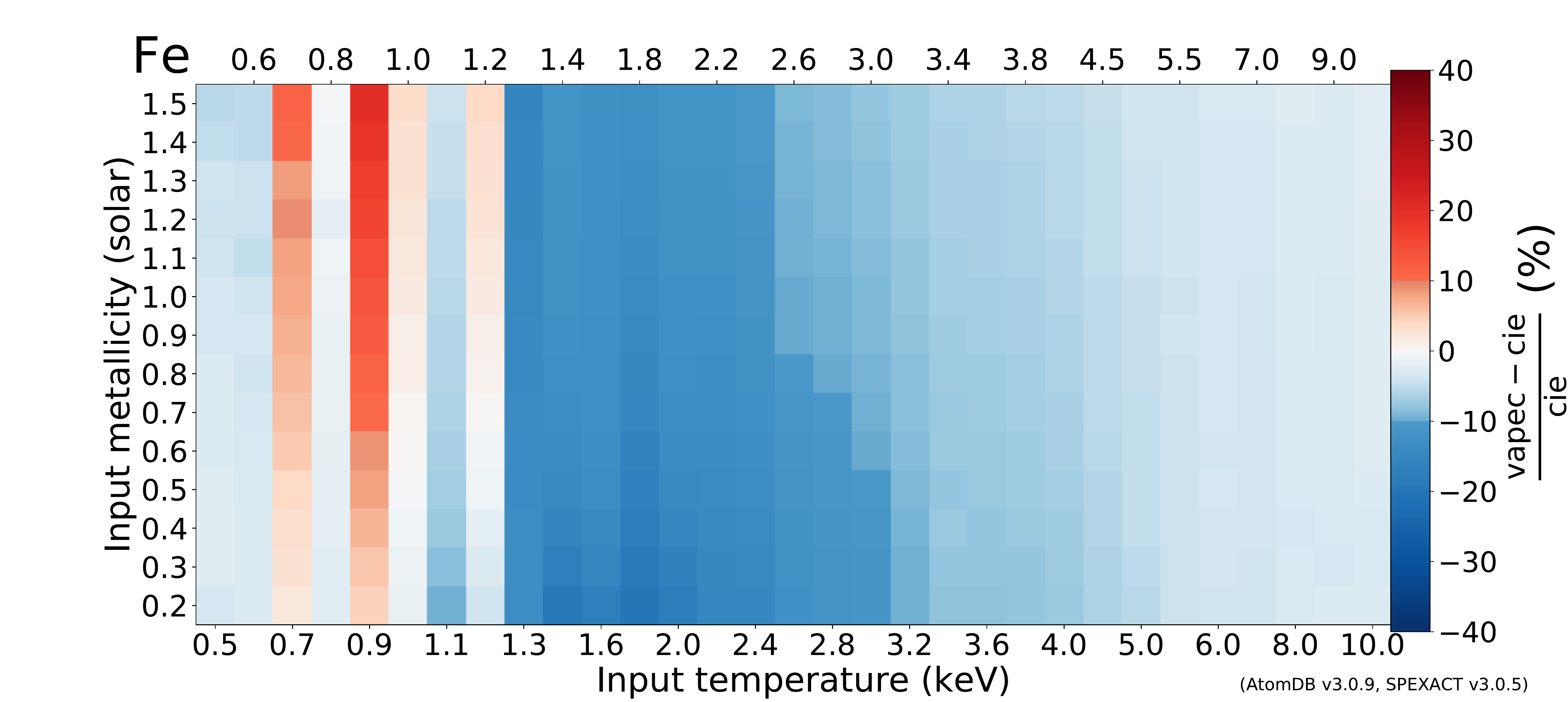} \\
\includegraphics[width=0.477\textwidth,trim={2.4cm 0 0 0},clip]{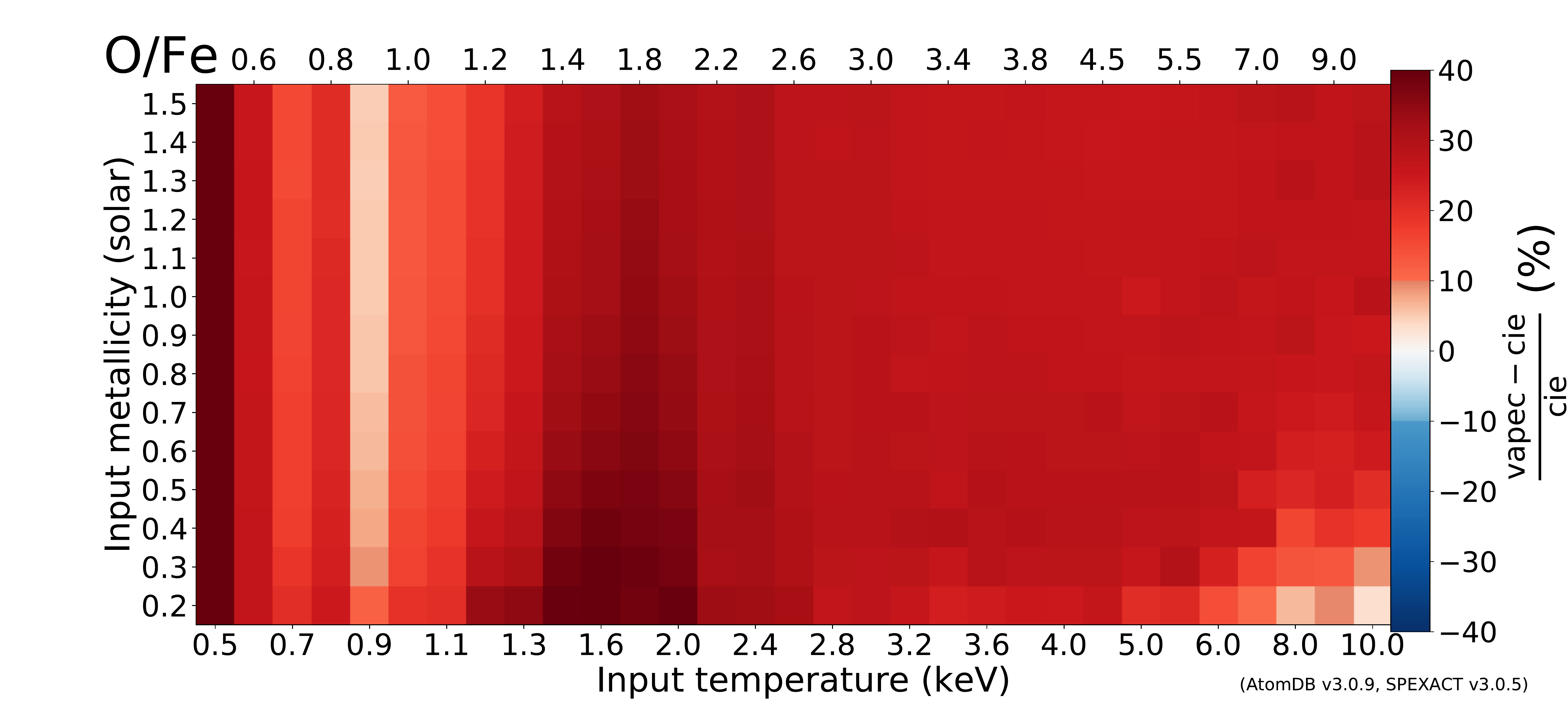}
\includegraphics[width=0.477\textwidth,trim={2.4cm 0 0 0},clip]{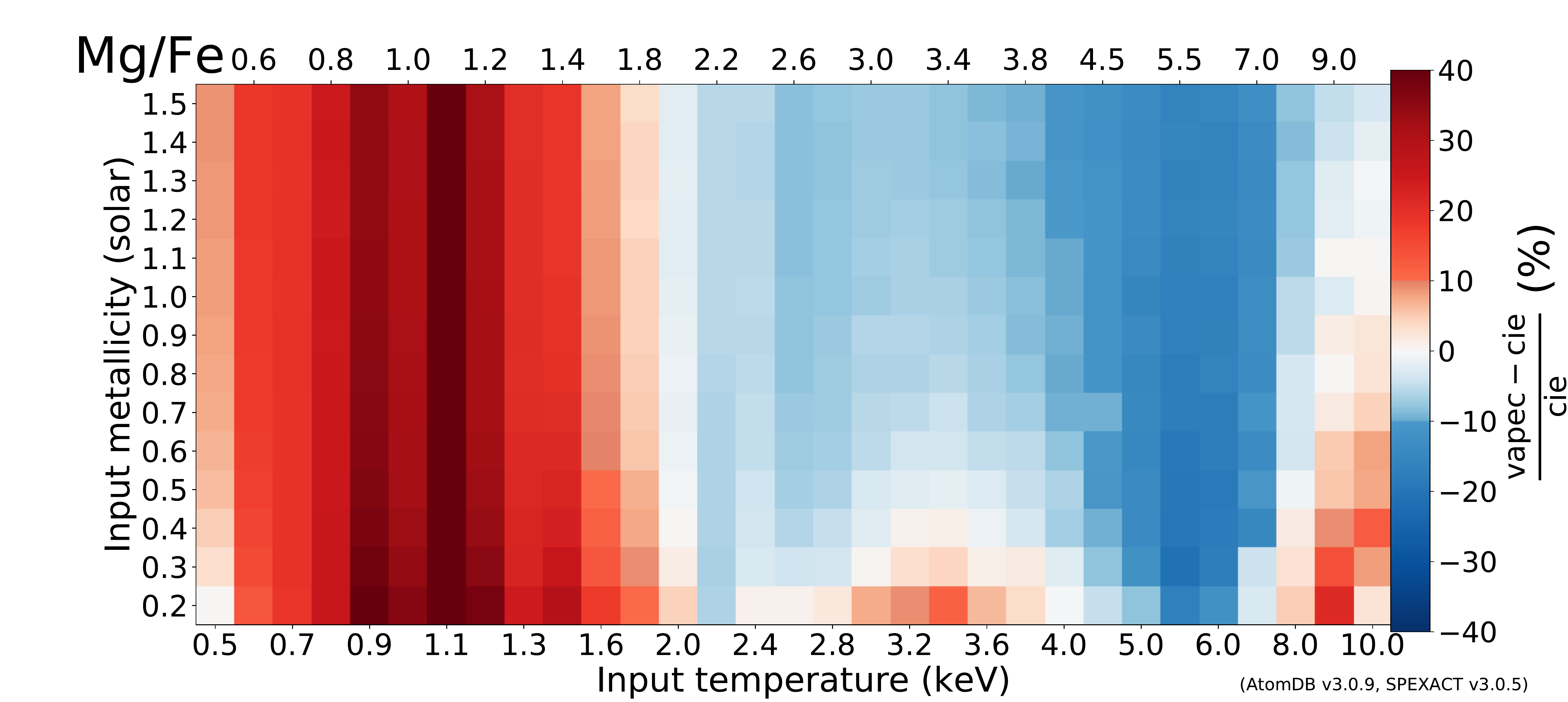} \\
\includegraphics[width=0.477\textwidth,trim={2.4cm 0 0 0},clip]{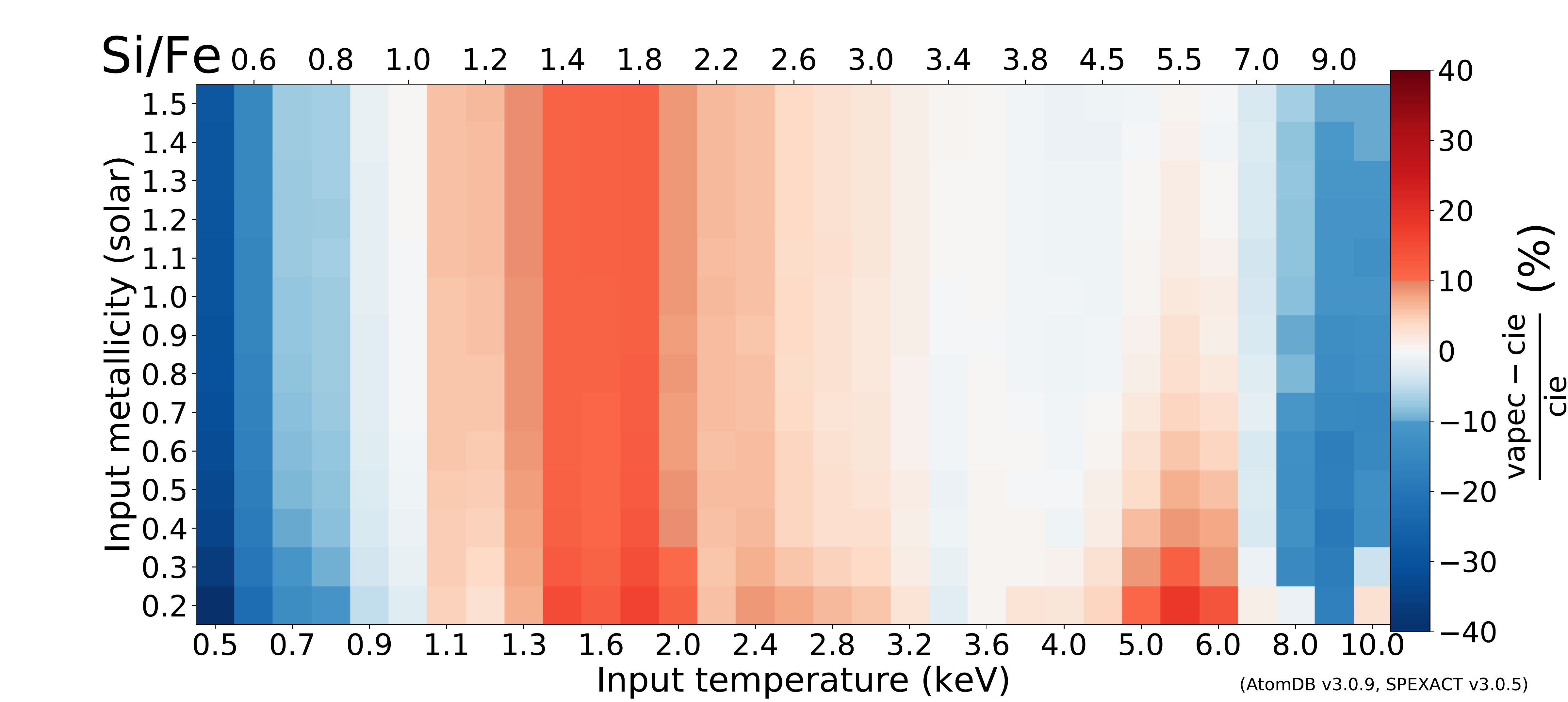}
\includegraphics[width=0.477\textwidth,trim={2.4cm 0 0 0},clip]{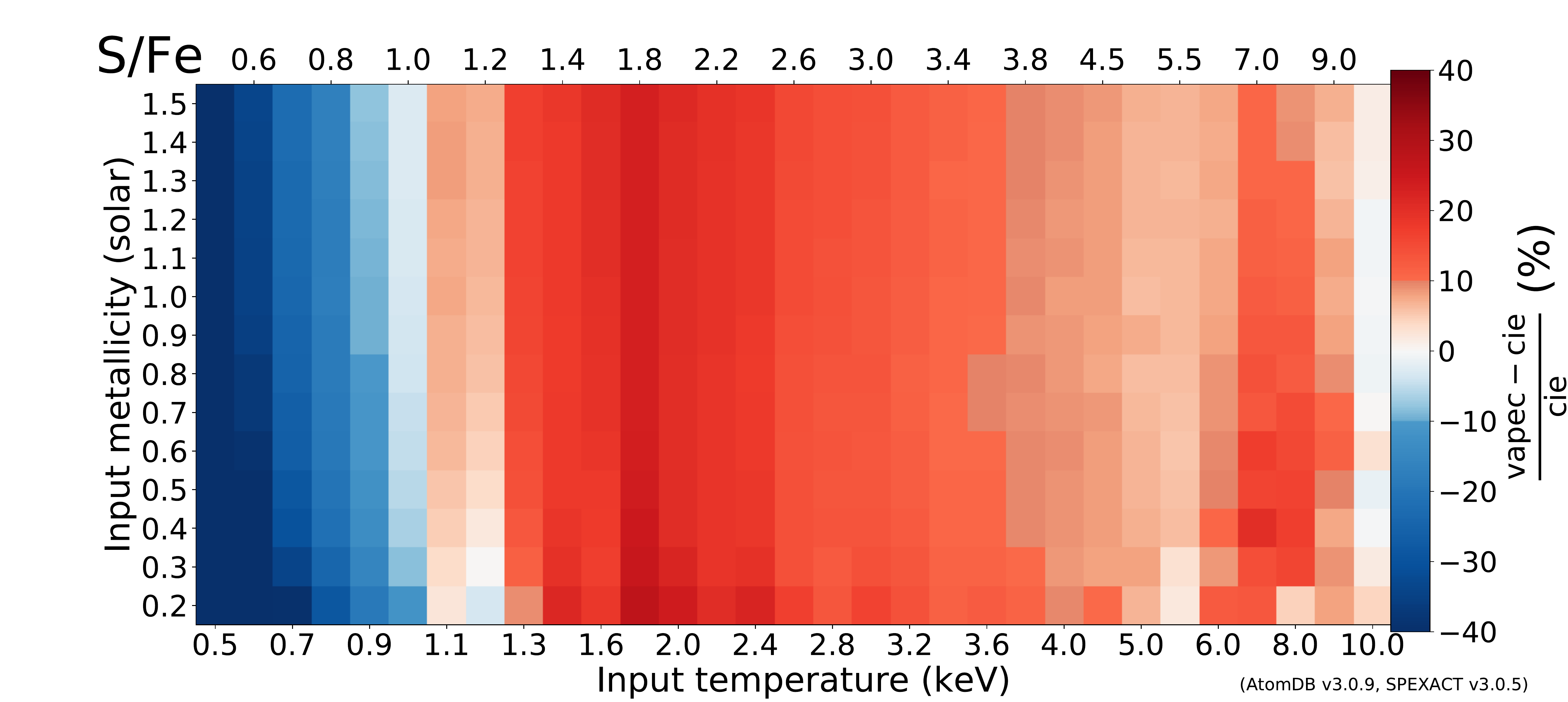} \\
\caption{\enspace Systematic temperature and abundance deviations of the \texttt{vapec} (AtomDB v3.0.9) model with respect to the \texttt{cie} (SPEXACT v3.0.5) model, for a grid of initial temperatures and metallicities (obtained with \textit{XMM-Newton}/MOS\,1 spectra; see text for details). Deviations beyond $\pm$10\% are marked with darker colors.\label{fig:apec_vs_SPEX}}
\end{figure*}

\section{SPEXACT vs. AtomDB}\label{sec:atomic_codes}

In this work, we aim to provide the community with a simple, though comprehensive set of quantified systematic uncertainties between the \texttt{cie} (SPEXACT v3.0.5) and the \texttt{vapec} (AtomDB v3.0.9) models in terms of temperature, Fe abundance (usually tracing the overall metallicity), and X/Fe abundance ratios, assuming plasmas with various temperature and chemical properties. Because the abundance reference tables of \citet{anders1989} are widely used in the literature (and remain the default option in XSPEC), we choose to refer to them in this work.

\subsection{Methodology}\label{sec:methodology}

As a first step, we use SPEX to simulate a series of redshifted ($z = 0.01$), absorbed ($n_\text{H} = 2 \times 10^{20}$ cm $^{-2}$) \texttt{cie} plasma models convolved by the \textit{XMM-Newton}/MOS\,1 instrumental response. In order to isolate the atomic code effects from other potential sources of systematic uncertainties, we restrict our exercise to the case of single-temperature plasma models with no (instrumental nor astrophysical) background (see also Sect.~\ref{sec:future}). The input temperature ($kT$) and Fe abundance parameters are selected from a grid of various values, spanning respectively within 0.5--10 keV and 0.2--1.5 solar. In each model, the abundances of other elements are tied to the input value of the Fe abundance. In order to get the exact count rate predicted by the models at each energy channel, the Poisson noise is set to zero in our simulations. 

The second step consists of fitting each of these \texttt{cie}-generated spectra with a single-temperature redshifted, absorbed \texttt{vapec} model. This can be done directly in SPEX by reading the AtomDB tables into the customisable \texttt{user} model via the \texttt{pyspextools}\footnote{\href{https://spex-xray.github.io/pyspextools}{https://spex-xray.github.io/pyspextools}} module. The fits are performed within the 0.5--10 keV band using C-statistics and the free parameters of the \texttt{vapec} model are the normalisation, the temperature, and the O, Mg, Si, S, and Fe abundances. Because they are known to be unresolved or undetectable in CCD-like spectra of low- or high-temperature plasmas (or in both), hence to be dominated by other sources of uncertainties, the abundances of the other elements (e.g. N, Ne, Ar, Ca) are left tied to Fe. The relative deviations between the \texttt{cie} input values of a given parameter and its corresponding \texttt{vapec} best-fit value can be then visualised on a grid containing all the initially assumed plasma temperature and Fe abundances (Fig.~\ref{fig:apec_vs_SPEX}). Additional tables including these numbers are provided separately\footnote{\href{https://github.com/mernier/SPEX_XSPEC}{https://github.com/mernier/SPEX\_XSPEC}}.

\subsection{Results \& Discussion}\label{sec:results}

Whereas the top left panel of Fig.~\ref{fig:apec_vs_SPEX} shows that the \texttt{cie} vs. \texttt{vapec} deviations in temperature are relatively limited ($\lesssim$10\% and $\lesssim$14\% only for 0.6 keV and 0.5 keV plasmas, respectively), it clearly appears that atomic code differences affect chemical abundances in a more significant way. 

\begin{figure}[t]
\centerline{\includegraphics[width=0.5\textwidth,trim={0.8cm 2.2cm 1.2cm 1.8cm},clip]{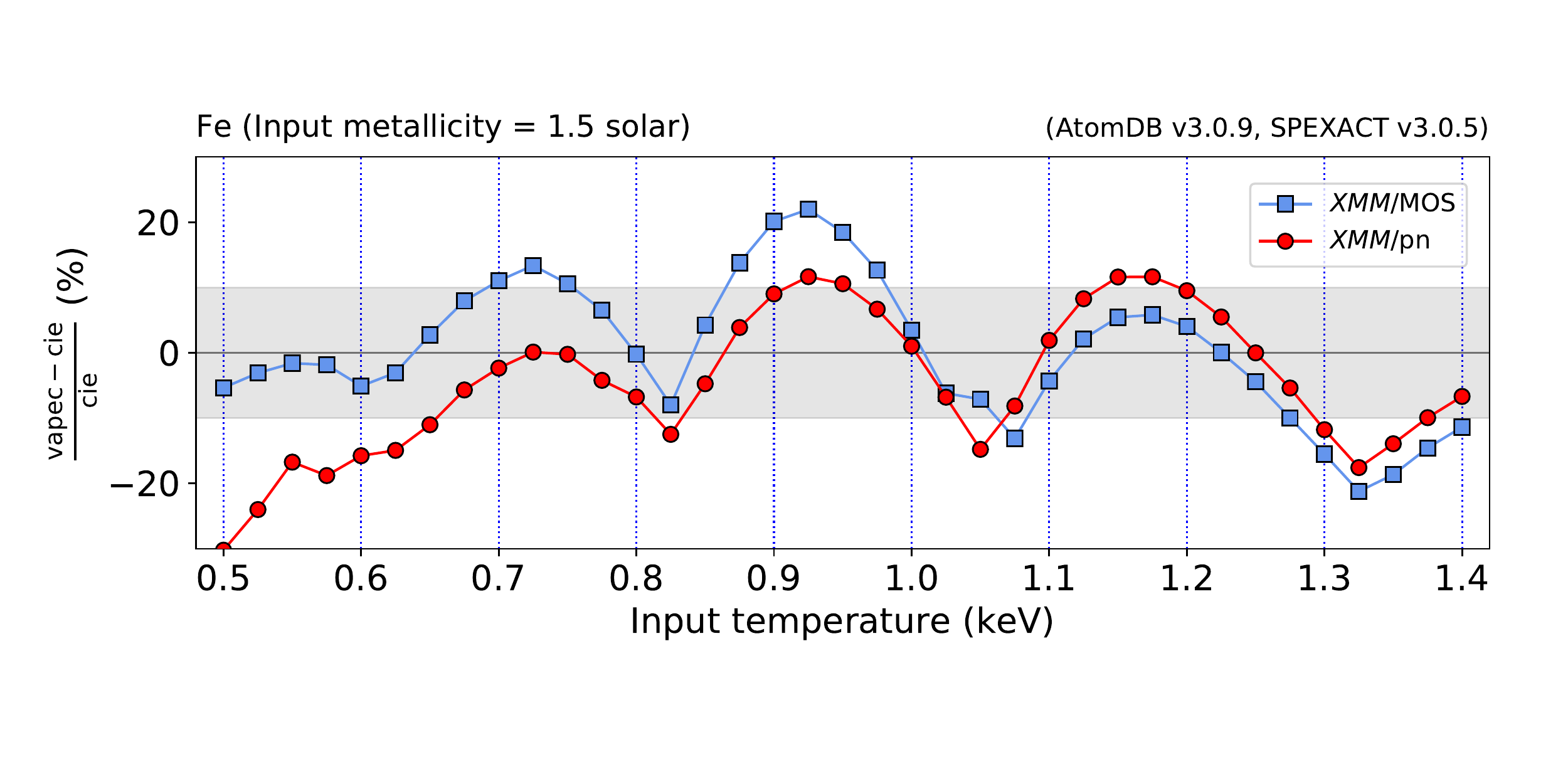}}
\caption{\enspace Systematic \texttt{vapec} vs. \texttt{cie} Fe deviations, for a finer grid of initial temperatures below 1.4 keV, at fixed input metallicity (1.5 solar). The blue dotted vertical lines refer to the lower grid resolution presented in Fig.~\ref{fig:apec_vs_SPEX}, while the grey area delimitates the $\pm$10$\%$ limits. \label{fig:zoom}}
\end{figure}

As shown in the top right panel, the Fe abundance is well recovered by \texttt{vapec} (<10\% discrepancies) for hot plasmas, i.e. above $\sim$3 keV. Beyond these temperatures, the Fe abundance is probed mainly via its K-shell transitions ($\sim$6.6 keV), which are now relatively well understood---especially after the data release of the SXS spectrum of Perseus \citep{THC2018_atomic}. Below these temperatures, however, Fe-L transitions start to take over, and many of them are modelled differently by SPEXACT and AtomDB. Because these lines are not resolved individually by moderate resolution instruments, the overall spectral shape of the Fe-L complex will appear slightly different from one atomic code to another. Due to the higher count rate of the Fe-L complex, the fits will be highly affected by this energy band. Consequently, slight differences in such spectral shapes may result into significant \texttt{cie} vs. \texttt{vapec} discrepancies. In fact, in intermediate temperature plasmas (1.3--3 keV), \texttt{vapec} systematically underestimates the Fe abundance by 10--20\% compared to \texttt{cie}. Below 1.3 keV plasmas, these discrepancies are contained between -10\% and +20\%, with discrete apparent variations between different initial temperatures. In order to explore these abrupt variations, we reprocess \texttt{vapec} fits of \texttt{cie} simulated spectra at fixed input Fe abundance (chosen here as 1.5 solar, i.e. where the variations are the highest) with a refined grid of input temperatures. The results, shown in Fig.~\ref{fig:zoom} (blue curve), reveal a complex structure of these \texttt{cie} vs. \texttt{vapec} deviations, with a series of smooth peaks and more abrupt drops as a function of the input temperature, thereby explaining the apparent discontinuous pattern seen in Fig.~\ref{fig:apec_vs_SPEX} (top right). The drops are the signature of the linear interpolation of \texttt{vapec} between its pre-calculated spectra (separated by $\log T = 0.10$). This will be corrected in a future version of the code (A. Foster, private comm.).

The four bottom panels of Fig.~\ref{fig:apec_vs_SPEX} show in a similar way \texttt{cie} vs. \texttt{vapec} deviations for the O/Fe, Mg/Fe, Si/Fe, and S/Fe abundance ratios. A noticeable case is the O/Fe ratio, for which the relative deviations span between +3\% and +45\% with no large dependency on the initial plasma conditions. The three other ratios show a finer temperature-dependent structure, with corresponding deviations ranging within $[-21\%, +42\%]$, $[-40\%, +18\%]$, and $[-80\%, +27\%]$ for the Mg/Fe, Si/Fe, and S/Fe ratios, respectively. The Si/Fe is clearly the most reliable ratio, as only plasmas cooler than 0.7 keV and hotter than 8 keV have discrepancies beyond $\pm$15\%. At cool (0.9--1 keV) and intermediate (3--4.5 keV) plasma temperatures, \texttt{cie} and \texttt{vapec} even match within less than 5\% for this ratio.

Another question of interest is whether atomic code uncertainties depend on the considered instrument. To check this possibility, we reprocess our spectral simulations and fits using the \textit{XMM-Newton}/pn instrumental response instead of the MOS\,1 one. Although no apparent modification of the output grid pattern is observed in any of the investigated parameters, we note slight but significant differences in the amplitude of the Fe deviations for cool plasmas. This is further illustrated in Fig.~\ref{fig:zoom} (i.e. based on finer grid of input temperatures), where the \texttt{cie} vs. \texttt{vapec} deviations obtained with pn (red curve) are often clearly distinct from those obtained with MOS (blue curve). This indicates that atomic code uncertainties reflect not only the intrinsic model-to-model discrepancies, but also propagate via their convolution with the instrumental response. In fact, different responses translate into different relative weights of the fit as a function of the energy (as some bands may appear more or less bright, hence with lower or higher error bars, respectively). After instrumental convolution, some parts of the Fe-L complex, containing critical lines that may be not well implemented yet, may be fitted with more or less priority.

Beyond raw measurements, also astrophysical interpretations may be significantly affected by all these code-related uncertainties. For instance, the slope of radial abundance profiles---crucial for inferring the ICM history and metal transport processes---may be code-dependent if the temperature gradient is important (e.g. in cool-core systems). In addition, further uncertainties on the chemical composition of the ICM (and on its relative SNIa/SNcc contribution) are worth considering and being quantified in future work.

\section{Future prospects}\label{sec:future}

In this work, we have provided a systematic comparison of inferred temperature and abundances between the \texttt{cie} (SPEXACT v3.0.5) and the \texttt{vapec} (AtomDB v3.0.9) models in the simple case of a single-temperature plasma seen through moderate (i.e. CCD-like) resolution spectroscopy. Despite the outstanding efforts that have been accomplished to greatly improve these two codes and make them converge \citep[especially thanks to the \textit{Hitomi} observations of Perseus;][]{THC2018_atomic}, abundance measurements still suffer from systematic uncertainties, in particular for cool plasmas (sometimes more than 20\% or even 30\%). Clearly, uncertainties related to the modelling of Fe-L transitions play a crucial role here, and future improvements of these two codes will certainly help to reduce these uncertainties. For instance, recently updated calculations on Fe-L transitions tend to revise upper the O/Fe ratio in the \texttt{cie} model \cite[e.g.][not used here]{gu2019}, potentially improving its agreement with \texttt{vapec}.

Admittedly, the comparison provided in this work is only a first step, and several important questions remain. For instance, how these atomic code uncertainties propagate with other biases that may affect the abundances (e.g. multi-temperature plasma, background uncertainties, etc.) has yet to be determined. In addition, the same exercise could be extended to other instrumental responses \cite[e.g. \textit{Chandra}/ACIS---see also][\textit{XRISM}/Resolve, \textit{Spektr-RG}/eROSITA, \textit{Athena}/X-IFU]{schellenberger2015}. Ultimately, comprehensive comparisons between these two codes should be tested on real observations in order to firmly assess potential astrophysical implications and their consequences on our knowledge of the ICM enrichment. This next step is left for future work (Lakhchaura et al., in prep).

Meanwhile, we hope that this basic attempt to quantify up-to-date atomic code uncertainties will be useful to the X-ray plasma community.


\section*{Acknowledgments}

We thank the anonymous referee as well as A. Foster and \'{A}. Bogd\'{a}n for useful discussions. This work was supported by the \fundingAgency{Lend\"ulet LP2016-11 grant} awarded by the \fundingNumber{Hungarian Academy of Sciences}. 
The SRON Netherlands Institute for Space Research is supported financially by \fundingNumber{NWO, the Netherlands Organisation for Scientific Research}. JM acknowledges the support from STFC (UK) through the University of Strathclyde \fundingAgency{UK APAP network grant} \fundingNumber{ST/R000743/1}.

%
%
%
%
%
%
%
%



\nocite{*}
\bibliography{Wiley-ASNA}%

\end{document}